# Experiences and attitudes toward working remotely from home in a time of pandemic: A snapshot from a New Zealand-based online survey

EDGAR PACHECO*

## Abstract

Due to the Covid-19 pandemic, employees from around the world were compelled to work remotely from home and, in many cases, without much preparation. A substantial body of international research has been conducted on the experiences and attitudes of remote workers as well as the implications of this phenomenon for organisations. While New Zealand research evidence is growing, most existing inquiry is qualitative. This paper provides a quantitative snapshot of remote working using survey data from participants whose jobs can be done from home (n=415). Data collection took place when the country was facing Covid-related measures.

Based on descriptive and inferential statistics, it was found that, not only was remote working common, but that hybrid working arrangements were also more prevalent. While half of the participants wanted to work from home more frequently, age, but not gender, was significantly associated with this preference. Another relevant finding is that perceived change in the workplace culture due to flexible work arrangements was significantly associated with preference for working remotely more often. Finally, the most common perceived barriers to working from home were slow internet speed, the need to attend face-to-face meetings, and limited space at home to work. The implications of the results are discussed and some directions for future research are proposed.

## Introduction

Just one week after the first Covid-19 case was reported in New Zealand on 28 February 2020, working remotely from home become an immediate necessity. Employees had to quickly adapt to the sudden change in work conditions not only to help prevent the spread of the virus but also to ensure business continuity. Before the pandemic, working from home was not a widespread practice in the country with only about a third of workers having ever done so (Stats NZ, 2019). In June 2020, during the lockdown and social distancing measures of levels 3 and 4 of the Covid-19

---

* Adjunct Research Fellow, School of Information Management, Victoria University of Wellington, New Zealand, and Senior Research Scientist (Social Sciences), WSP New Zealand.
Email: edgar.pacheco@vuw.ac.nz.

The author would like to thank InternetNZ (https://internetnz.nz/) for providing access to the raw data underpinning this article.

alert system (see Department of the Prime Minister and Cabinet, 2024), more than 40 per cent of employed people were reported to have done at least some of their work from home (Stats NZ, 2020). Similarly, a report by O'Kane et al. (2020) indicates that, prior to the Covid-19 pandemic, 74 per cent of employees surveyed had no regular experience of working from home, with 38 per cent reporting of never having done so. However, in May 2020, when the survey was conducted, 92 per cent of participants said they were working from home full time (O'Kane et al., 2020). Three years later, in 2023, a report on remote working in English-speaking countries, including New Zealand, found that full-time employees work on average 1.4 full paid days per week from home (Aksoy et al., 2023). Clearly, due to Covid-19, flexible remote work has become, to some extent, the 'new normal' for many in New Zealand's workforce.

Remote working refers to "working from home or another location outside an office [company premises or production facilities] at any time, which involves the increasing use of technology enabling workers to communicate with their workplace and supporting flexible working practices" (Battisti et al., 2022, p. 39). In the literature, the term 'remote working' is used interchangeably with 'working from home' and 'telework' (Green et al., 2020). We do so here, too, when discussing conceptual differences (e.g., Vartiainen, 2021) is outside the scope of this paper.

*Remote working in conventional conditions*

Before the pandemic, evidence about remote working was mostly produced in the context of voluntary or consensual arrangements; for instance, work time shared between the office and home (Platts et al., 2022). This 'conventional' remote working (Blahopoulou et al., 2022) was thought to provide employees with flexibility to manage challenges between work and life roles. In doing so, employees were allowed to work from home, reduced work hours or start work earlier/later (Bentley et al., 2016). Drivers for the growth of remote working before the pandemic were the rapid development of new technologies (e.g., broadband internet, social media, cloud computing, and networking tools), the obligation in some countries for firms to consider employee requests to move to flexible working arrangements, and proponents' claims that remote working arrangements may modernise workplace practices (Bentley et al., 2016). A number of studies have proposed that remote working is detrimental to employees' social, behavioural, and physical wellbeing (Eddleston & Mulki, 2017; Oakman et al., 2020). However, two meta-analysis studies (Gajendran & Harrison, 2007; Harker Martin & MacDonnell, 2012) about remote working under conventional or business-as-usual conditions show otherwise. Gajendran and Harrison (2007) examined 46 studies and found that remote working reduces work-family conflict, and it is associated with higher perceptions of autonomy and quality of employee-supervisor relationship. The benefits of remote working increase for those who work from home on a regular basis (Gajendran & Harrison, 2007). Another meta-analysis of 32 studies (Harker Martin & MacDonnell, 2012) found a small but positive relationship between remote working and organisational outcomes, and perceived benefits, such as increased productivity, more secure retention, strengthened organisational commitment, and improved performance (Harker Martin & MacDonnell, 2012).

In New Zealand, remote working has also been a matter of research and discussion for quite some time. In the context of business-as-usual conditions, Schoeffel et al. (1993) conducted in the early 1990s a small-scale survey to examine the prevalence of remote working, aligning their findings with international trends. Similarly, Bentley and Yoong (2000) undertook an interview-based study to explore the experiences of knowledge workers from two consultancy organisations. Meanwhile, a study by Bentley et al., (2013) delved into the relationship between remote working and productivity, work satisfaction, and work performance among managers and employees from New Zealand and Australia. Bentley (2016) also studied the impact of organisational support on remote workers' wellbeing. Nevertheless, not all research in New Zealand concerning remote work was carried out under business-as-usual conditions. The disruption caused by the Christchurch earthquakes between 2010 and 2011 prompted a couple of studies of remote work in the context of a natural disaster. In this regard, a case study by Green at al. (2017) looked at the impacts of remote working on business continuity and employee wellbeing. In the same context, Donnelly and Proctor-Thomson (2015) studied the home working experiences of public sector employees and their managers after the earthquakes. Consequently, remote working has garnered interest in New Zealand, with research offering valuable insights for both the private and public sectors not only during conventional circumstances but also in time of natural disasters.

*Remote working in the Covid-19 era*

With the COVID-19 pandemic, however, remote working shifted from discretionary flexible work policies to a 'forced flexibility' (Jogulu et al., 2023). Under a challenging context with frequent lockdowns causing work disruptions, organisations and employees had to quickly adapt to new mandatory and full-time requirements and arrangements with little or no prior preparation (Jogulu et al., 2023). The speed and widespread shift to this new style of working without considerable organisational experience or awareness of its complications, created potential for unforeseen challenges (Chow et al., 2022). In addition to limited organisational preparedness, technological limitations and/or managerial reluctance were also challenges for the implementation of remote working both before and during the pandemic (Green et al., 2020).

Several overseas studies have investigated the outcomes of remote working for employees during the Covid-19 pandemic but with conflicting results. A study based on data from 29 European countries by Ipsen et al., (2021) found that, during lockdowns, remote workers reported more positive experiences (i.e., work-life balance, improved work efficiency, and greater work control) than negative ones. In Spain, a study reported that satisfaction with remote working arrangements was linked with positive direct effects on subjective wellbeing and self-reported work performance (Blahopoulou et al., 2022). In India, Mohammed et al. (2022) found that job satisfaction was positively related to perceived work autonomy among employees working from home. On the other hand, some evidence points out the disadvantages of remote working. For instance, a South African survey-based study (Matli, 2020) found that first-time remote workers, those who had never worked from home, reported work overload and pressures to perform in a timely manner. Meanwhile in Colombia, Camacho and Barrios (2022) found that work overload and conflicts between work and home boundaries impacted negatively stress levels, thus diminishing employees' job satisfaction. Similarly, a study conducted in Austria looking at the accumulation of stress among remote workers found a reduction in wellbeing, perceived productivity, and engagement in the first 11 months of the pandemic (Straus et al., 2022). Meanwhile the role of social isolation because of lockdown among a group of Italian remote workers was explored by Toscano and Zappalà (2020), who found that it negatively affected overall satisfaction with remote working. The reasons behind the varying results in the literature remain ambiguous, however the specific context of each jurisdiction, and the extent of the impact of the pandemic on them, might offer some explanation. The complexity of the situation suggests that multiple variables are at play, creating a challenging landscape for researchers, organisations, and policy makers to navigate and understand thoroughly.

While the contrasting evidence regarding the outcomes of remote working mostly derives from the first year of the pandemic, recent data from the Pew Research Center (Parker, 2023) suggest that working from home, nowadays, might be a choice rather than a necessity for most remote workers in the United States as 61 per cent said they have chosen not to go to their workplace.

The transition in New Zealand from conventional remote work practices to the realm of a forced flexibility driven by the necessity of adapting to the challenges posed by the Covid-19 pandemic has also given rise to a growing body of research. Studies have delved into the advantages and drawbacks of remote work, its effects on aspects such as wellbeing and work-life balance, as well as its influence on organisational culture (Bradley et al., 2023; Gorjifard & Crawford, 2021; Green et al., 2020), public service provision (Lips & Eppel, 2022), its representation in New Zealand media (Jamindar & Bone, 2023), and the impact on employee surveillance (Blumenfeld et al., 2020). Despite this growing body of Covid-related evidence, most New Zealand-based research has relied on qualitative inquiry, which is highly contextual. Only publications by O'Kane et al. (2020), Stats NZ (Stats NZ, 2020) and, more recently, Hedges et al. (2023) have used quantitative data. While relevant, these quantitative studies only provide broad trends about remote working (see the beginning of this section for details). Associations between variables were not tested, nor was analysis conducted on key demographics, such as age or gender, thereby indicating a discernible knowledge gap. This situation contrasts with Australia where several quantitative studies have explored remote working during and after Covid-19 (see Baxter & Warren, 2021; Cornell et al., 2022; Gallagher, 2021; Marzban et al., 2021; Oakman et al., 2022; Petrie, 2023).

*The current study*

In light of the above, the current quantitative work seeks to provide New Zealand-based evidence about employees' experiences and attitudes toward working remotely from home in the context of the Covid-19 pandemic. Based on data from a larger survey, the current study explores different aspects related to remote working, such as how often participants work remotely from home, their perceptions regarding doing so more often, perceived change in the workplace's culture, and perceived barriers to working from home. Furthermore, the purpose of the current work is to identify statistical associations of these aspects with key demographics, namely gender, age, and region, as well as the frequency of working remotely. We believe the evidence from this paper will contribute to better understanding remote working in the New Zealand context and inform the implementation or improvement of policies and practices supporting flexible work arrangements.

## Method

*Study design*

Data for this paper came from *New Zealand's Internet Insights*, a modular online panel survey run by InternetNZ. The survey was designed to gather self-reported data about different aspects of people's interaction with digital technologies, remote working being one of them. The survey's purpose was to describe general trends in digital behaviours by organising and summarising data. However, making statistical inferences based on associations found in the sample was not the goal of InternetNZ's survey. As no further analysis was performed, the researcher asked the organisation for the anonymised raw survey data and data dictionary to apply additional analysis via inferential statistics with the purpose of determining statistically significant group differences, namely gender, age, and region. InternetNZ provided the anonymised raw survey data and data dictionary to the researcher on request, allowing secondary data analysis for the survey to be applied.

Secondary data analysis is, as described by Vartanian (2010), the analysis of any data in order to answer a research question(s) different from the questions(s) for which the data was originally gathered. The benefits of this research approach being that researchers can conduct inquiries when time and resources are limited or when research activities are disrupted due to unexpected events (Spurlock, 2020), for instance the Covid-19 pandemic (Pacheco, 2022, 2024).

The full survey was administered online. Online surveys are cost-effective, easier to administer (Sue & Ritter, 2012) and are increasingly used in social and policy research (Lehdonvirta et al., 2021). In terms of exploratory inquiry, online surveys have been shown to be a useful technique for gathering evidence about people's access, activities, and attitudes regarding the digital environment (Pacheco & Melhuish, 2019; 2020).

The sample for the full survey comprised 1,001 New Zealanders aged 18 and older. Data collection was conducted by market research company, Kantar Public, on behalf of InternetNZ. Using its online consumer panels, Kanter Public employed a combination of pre-survey quotas and post-survey weighting using census-based estimates to ensure results are representative of all New Zealanders by key demographics, such as age, gender, and region.

The survey questionnaire underwent cognitive testing and pilot testing to aid completion by participants. Consent to take part in the study was obtained from all participants who had the right to withdraw from the research at any time. The maximum margin of error is +3.1 per cent at the 95 per cent confidence interval.

*Data collection context*

Data gathering was carried out from 3 to 17 November 2021. It is necessary to note that, when data collection began, parts of the country (Auckland, Upper Northland, and Waikato) were at Level 3 (Restrict) of New Zealand's Covid-19 alert system while the rest were on Level 2 (Reduce).

- At Level 3 (Restrict), there was medium risk of community transmission, with active but managed clusters, meaning people in Auckland, Upper Northland, and Waikato faced some restrictions but not lockdown. They had to stay at home and keep their 'bubble' small, with gatherings of up to 10 people allowed. It was obligatory to wear a face mask in some settings. When in public, a two-metre distance from others had to be maintained. Long distance travel was restricted and working from home was recommended; only people unable to do so were permitted to return to businesses that could safely open under Level 3. Public facilities remained closed while early childhood centres and schools were open for students up to Year 10 who could not learn from home.

- Meanwhile, Level 2 (Reduce) meant a low risk of community transmission although there could be limited community transmission and active clusters in more than one region. Prevention measures were less restrictive in Level 2. People were allowed to socialise in groups and connect with friends and their whānau (family) in person, and gatherings of up to 100 people being permitted in a defined setting (e.g., weddings). People were also allowed to return to their workplace, but alternative ways of working were encouraged. Business and services could open but additional health measures needed to be in place (see Department of the Prime Minister and Cabinet, 2024).

*Study sample*

Of the participants who undertook the full survey (n=1,001), 674 indicated that they were in paid employment and could access the internet from home. Then, this group of 674 participants was asked whether they do a type of work that allows them to work from home as well as from their workplace. A total of 415 participants responded positively meeting the minimum inclusion criteria for the final sample of the study. Table 1 breaks down the characteristics of the sample according to demographic variables.

**Table 1. Sample Demographic Distribution**

| Demographic | Number | % |
|---|---|---|
| **Age group** | | |
| 18-29 | 95 | 22.9 |
| 30-49 | 183 | 44.1 |
| 50-64 | 109 | 26.3 |
| 65 and over | 28 | 6.7 |
| **Gender** | | |
| Male | 198 | 47.7 |
| Female | 217 | 52.3 |
| **Region** | | |
| Auckland | 163 | 39.3 |
| Wellington | 49 | 11.8 |
| Canterbury | 45 | 10.8 |
| Rest of North Island | 125 | 30.1 |
| Rest of South Island | 33 | 8.0 |
| **Total** | 415 | 100.0 |

**Measures**

*Demographics*

Participants provided demographic information regarding their gender, age group, and the region where they live. Regarding gender, the categories male; female; and gender diverse were provided. None of them identified as gender diverse. In addition, and for analysis, participants were aggregated into four age group categories: 18-29 years; 30-49 years; 50-64 years; and 65 years and over. In terms of region, their responses were aggregated as follows: Auckland; Wellington; Canterbury; rest of North Island; and rest of South Island.

*Working from home*
Participants were asked *Do you work remotely from home?* with response options being:
Yes – I do it all the time; Yes – I do it sometimes; Yes – I've done it once or twice; and No – I've never worked remotely from home.

*Preferences for working from home more frequently*
Participants were asked *Would you like to work from home more frequently than you currently do?* with the following response options: Yes; No; and Don't know.

*Perceived impact of working from home on the organisational culture*
In summary, organisational culture refers to a set of values that guides the activities of an organisation's members (Vandenberghe & Peiro, 1999). The following question was asked of participants: *Thinking of your workplace, how has the culture been affected by more people working from home over the last year and a half?* The following options were provided: It's better; The workplace culture hasn't changed; It's worse; Not applicable, there's been no change in the number of people working from home; and Don't know.

*Perceived barriers to working from home*
A multi-response question asked the following of those participants who previously said answered that they would like to work from home more frequently (n=210): *Which of the below are barriers for you working from home more often?* The options included: My laptop or mobile phone isn't good enough; 'My internet speed isn't fast enough; I don't have a space at home to work; I have too many face-to-face meetings, I need to be at work; My employer doesn't offer the technological support for me to work from home; My employer doesn't offer flexible working options; My employer doesn't encourage or frowns upon people working from home; and Other.

### ***Data analysis***
Categorical variables are presented in contingency tables as frequencies and percentages. Descriptive analysis and bivariate statistical inference were performed. To test the significance of association between two nominal variables, the Chi-square test of independence was employed. The Chi-square test of independence is widely recognised for its robustness in statistical analysis for examining the association between categorical variables (McHugh, 2013). It is important to note, however, that this test does not serve the purpose of establishing a causal relationship between two variables or determining the direction of influence between them. Despite this, since the survey variables are categorical, the Chi-square test was the best analytical choice. Statistical significance was determined at $p < 0.05$. Cramer's V was given as an effect size of the Chi-square test. Rea and Parker's (2014) interpretation was used to observe the strength of association. In this respect, a Cramer's V below 0.10 means a negligible association, between 0.10 and below 0.20 indicates a weak association, between 0.20 and below 0.40 represents a moderate association, and between 0.40 and 0.60 indicates a relatively strong association. Categories regarding age groups and region of residence were collapsed into larger categories to facilitate analysis of the observed counts. All analyses were performed using the Jamovi software, version 2.3 (The jamovi project, 2023).

## Results

Firstly, the association of how often participants work remotely from home with each of the three demographic variables (i.e., gender, age, and region) was tested. As can be seen by the frequencies cross-tabulated in Table 2, there was a significant association between remote working and region, $\chi^2$ (12, N = 415) = 31.8, p = .002. However, the effect size for this finding, Cramer's V, was weak, .16. Participants from the Auckland region were more likely to work from home all the time compared to those from the rest of the country. In contrast, those from the Wellington region were more likely to work from home sometimes. Regarding remote working and gender, a Chi-square test of independence showed that there was no significant association between the variables, $\chi^2$ (3, N = 415) = 5.95, p = .114. Similarly, no statistical significance was found for remote working by age, $\chi^2$ (9, N = 415) = 15.3, p = .084.

**Table 2. Frequency of Working from Home by Gender, Age, and Region**

| | Working remotely from home | | | | | | | | | | |
|---|---|---|---|---|---|---|---|---|---|---|---|
| | All the time | | Sometimes | | Once or twice | | Never | | Total | | *p* |
| Demographics | N | % | n | % | n | % | n | % | n | % | |
| Male | 74 | 37.4 | 98 | 49.5 | 22 | 11.1 | 4 | 2.0 | 198 | 100.0 | .114 |
| Female | 78 | 35.9 | 93 | 42.9 | 34 | 15.7 | 12 | 5.5 | 217 | 100.0 | |
| 18-29 | 29 | 30.5 | 39 | 41.1 | 19 | 20.0 | 8 | 8.4 | 95 | 100.0 | .084 |
| 30-49 | 66 | 36.1 | 89 | 48.6 | 24 | 13.1 | 4 | 2.2 | 183 | 100.0 | |
| 50-64 | 44 | 40.4 | 50 | 45.9 | 12 | 11.0 | 3 | 2.8 | 109 | 100.0 | |
| 65 and over | 13 | 46.4 | 13 | 46.4 | 1 | 3.6 | 1 | 3.6 | 28 | 100.0 | |
| Auckland | 70 | 42.9 | 77 | 47.2 | 11 | 6.7 | 5 | 3.1 | 163 | 100.0 | .002 |
| Wellington | 16 | 32.7 | 29 | 59.2 | 4 | 8.2 | 0 | 0.0 | 49 | 100.0 | |
| Canterbury | 14 | 31.1 | 20 | 44.4 | 11 | 24.4 | 0 | 0.0 | 45 | 100.0 | |
| Rest of North Island | 38 | 30.4 | 51 | 40.8 | 27 | 21.6 | 9 | 7.2 | 125 | 100.0 | |
| Rest of South Island | 14 | 42.4 | 14 | 42.4 | 3 | 9.1 | 2 | 6.1 | 33 | 100.0 | |
| Total | 152 | 36.6 | 191 | 46.0 | 56 | 13.5 | 16 | 3.9 | 415 | 100.0 | |

We also looked at the association of preferences for working remotely from home more frequently with gender, age, and region. In terms of age, a significant association was found, $\chi^2$ (3, N = 382) = 20.5, p < .001. The effect size for this result, Cramer's V, was moderate, .23. Participants aged 18-29 years were more likely to say that they would like to work from home more frequently. Similarly, a significant but weak association was found in terms of region, $\chi^2$ (4, N = 382) = 9.94, p = .041, V = .16. Participants from the Auckland region were more likely to say they would like to work from home more frequently. Regarding gender, no significant difference was found, $\chi^2$ (1, N = 382) = 0.00521, p = .942. See Table 3 for details.

**Table 3. Preference for Working from Home More Frequently by Gender, Age, and Region**

| Demographics | Would like to work from home more frequently * | | | | Total | | *p* |
|---|---|---|---|---|---|---|---|
| | Yes | | No | | | | |
| | n | % | n | % | n | % | |
| Male | 103 | 54.8 | 85 | 45.2 | 188 | 100.0 | .942 |
| Female | 107 | 55.2 | 87 | 44.8 | 194 | 100.0 | |
| 18-29 | 60 | 66.7 | 30 | 33.3 | 90 | 100.0 | <.001 |
| 30-49 | 100 | 59.2 | 69 | 40.8 | 169 | 100.0 | |
| 50-64 | 45 | 45.0 | 55 | 55.0 | 100 | 100.0 | |
| 65 and over | 5 | 21.7 | 18 | 78.3 | 23 | 100.0 | |
| Auckland | 97 | 63.0 | 57 | 37.0 | 154 | 100.0 | .041 |
| Wellington | 28 | 60.9 | 18 | 39.1 | 46 | 100.0 | |
| Canterbury | 22 | 50.0 | 22 | 50.0 | 44 | 100.0 | |
| Rest of North Island | 51 | 45.9 | 60 | 54.1 | 111 | 100.0 | |
| Rest of South Island | 12 | 44.4 | 15 | 55.6 | 27 | 100.0 | |
| Total | 210 | 55.0 | 172 | 45.0 | 382 | 100.0 | |

*Note: * Don't know (n=33) not included in analysis.*

On the other hand, when analysing changes in the work culture, the Chi-square test of independence did not find a significant association with any of the three demographic variables included in the study. In the case of gender, the results were $\chi^2$ (3, N = 370) = 4.71, p = .194. No significant association was found in terms of age, $\chi^2$ (9, N = 370) = 15.2, p = .086. Similarly, regarding region, the overall Chi-square test did not reach significance, $\chi^2$ (12, N = 370) = 19.7, p = .073. See Table 4.

**Table 4. Perceived Change in the Workplace Culture by Gender, Age, and Region**

| Demographics | Whether work culture has changed* | | | | | | | | Total | | *p* |
|---|---|---|---|---|---|---|---|---|---|---|---|
| | Better | | Hasn't changed | | Worse | | Not applicable | | | | |
| | n | % | n | % | n | % | n | % | n | % | |
| Male | 42 | 23.2 | 96 | 53.0 | 29 | 16.0 | 14 | 7.7 | 181 | 100.0 | .194 |
| Female | 43 | 22.8 | 84 | 44.4 | 37 | 19.6 | 25 | 13.2 | 189 | 100.0 | |
| 18-29 | 21 | 25.3 | 48 | 57.8 | 9 | 10.8 | 5 | 6.0 | 83 | 100.0 | .086 |
| 30-49 | 45 | 26.8 | 76 | 45.2 | 32 | 19.0 | 15 | 8.9 | 168 | 100.0 | |
| 50-64 | 16 | 16.3 | 45 | 45.9 | 22 | 22.4 | 15 | 15.3 | 98 | 100.0 | |
| 65 and over | 3 | 14.3 | 11 | 52.4 | 3 | 14.3 | 4 | 19.0 | 21 | 100.0 | |
| Auckland | 38 | 26.4 | 73 | 50.7 | 25 | 17.4 | 8 | 5.6 | 144 | 100.0 | .073 |
| Wellington | 12 | 27.3 | 23 | 52.3 | 7 | 15.9 | 2 | 4.5 | 44 | 100.0 | |
| Canterbury | 10 | 23.8 | 21 | 50.0 | 3 | 7.1 | 8 | 19.0 | 42 | 100.0 | |
| Rest of North Island | 19 | 16.8 | 51 | 45.1 | 27 | 23.9 | 16 | 14.2 | 113 | 100.0 | |
| Rest of South Island | 6 | 22.2 | 12 | 44.4 | 4 | 14.8 | 5 | 18.5 | 27 | 100.0 | |
| Total | 85 | 23.0 | 180 | 48.6 | 66 | 17.8 | 39 | 10.5 | 370 | 100.0 | |

*Note: * Don't know (n=45) not included in analysis.*

We also looked at preferences for working remotely from home more frequently and its association with changes in the work culture (see Table 5). A significant association was found between these two nominal variables, $\chi^2$ (3, N = 347) = 15.8, p = .001. The point estimate for effect size, Cramer's V, indicated a medium effect, .21. Participants who indicated that the work culture is better were more likely to say they would like to work from home more frequently.

**Table 5. Preference for Working from Home More Frequently and Perceived Change in the Workplace Culture**

| | Would like to work from home more frequently | | | | | | |
|---|---|---|---|---|---|---|---|
| Whether work culture has changed | Yes | | No | | Total | | *p* |
| | n | % | n | % | n | % | |
| It's better | 58 | 70.7 | 24 | 29.3 | 82 | 100.0 | .001 |
| It hasn't changed | 95 | 55.9 | 75 | 44.1 | 170 | 100.0 | |
| It's worse | 23 | 39.7 | 35 | 60.3 | 58 | 100.0 | |
| Not applicable, there's been no change in the number of people working from home | 16 | 43.2 | 21 | 56.8 | 37 | 100.0 | |
| Total | 192 | 55.3 | 155 | 44.7 | 347 | 100.0 | |

*Note: 'Don't know' not included in analysis.*

On the other hand, as shown in Table 6, when analysing how often participants work from home and their preferences for doing so more frequently, no significant association was found, $\chi^2$ (3, N = 382) = 2.54, p = .469.

**Table 6. Preference for Working from Home More Frequently and Frequency of Working from Home**

| Working from home | Would like to work from home more frequently * | | | | | | |
|---|---|---|---|---|---|---|---|
| | Yes | | No | | Total | | *p* |
| | n | % | n | % | n | % | |
| All the time | 72 | 51.4 | 68 | 48.6 | 140 | 100.0 | .469 |
| Sometimes | 107 | 59.1 | 74 | 40.9 | 181 | 100.0 | |
| Only once or twice | 25 | 52.1 | 23 | 47.9 | 48 | 100.0 | |
| Never worked from home | 6 | 46.2 | 7 | 53.8 | 13 | 100.0 | |
| Total | 210 | 55.0 | 172 | 45.0 | 382 | 100.0 | |

Finally, shown in Figure 1, a descriptive analysis of the results related to perceived barriers to working from home more often shows that the quality of internet speed was the most common issue among the participants who would like to work remotely from home more often (n=210). Other common barriers were the need to be at the workplace due to the quantity of face-to-face meetings, and the limited space at home to work remotely.

**Figure 1. Perceived Barriers to Working from Home More Often**

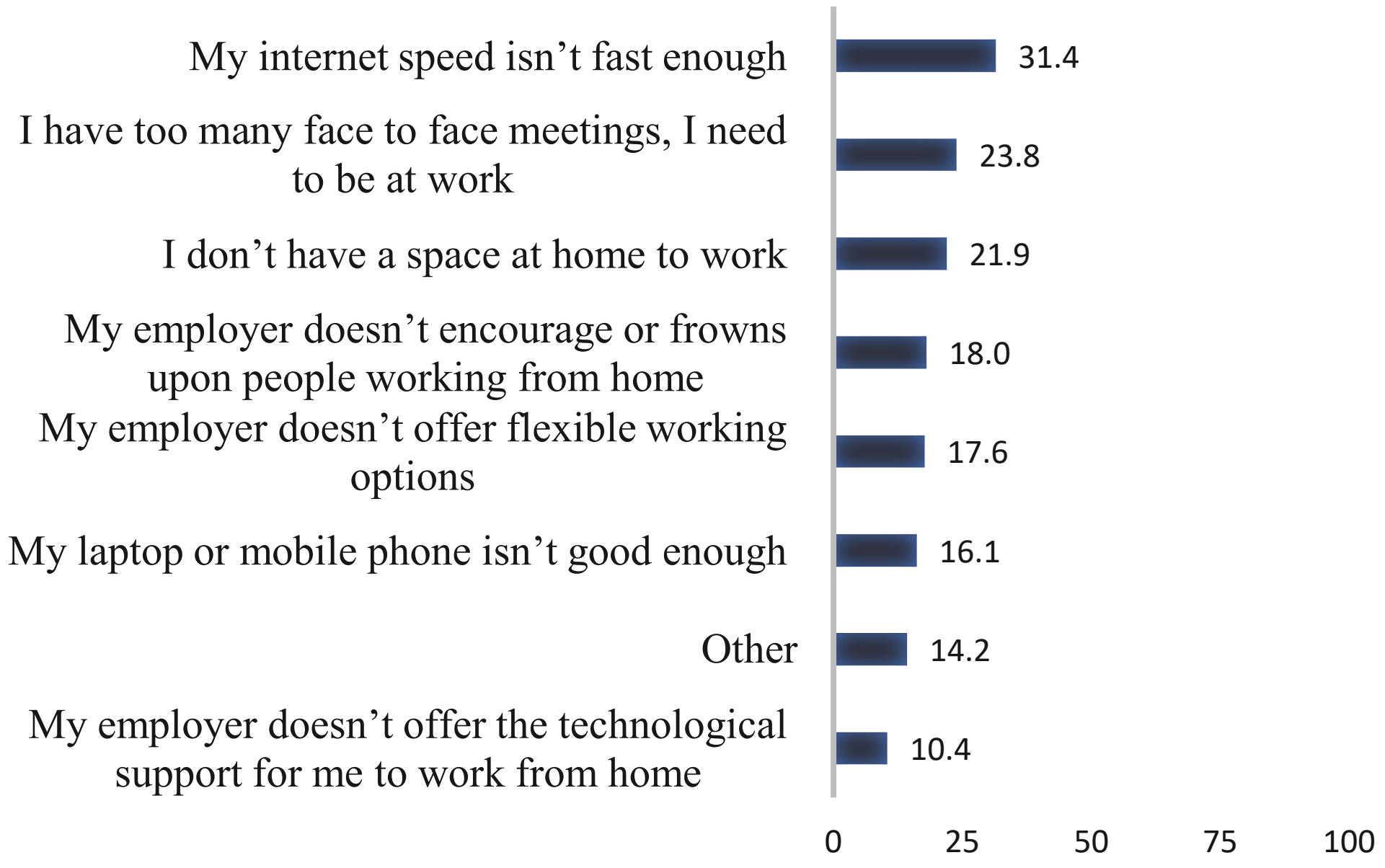

*Note: n=210. Multi-response question.*

## Discussion and conclusion

Working from home is a practice with which most participants in our sample were familiar as a consequence of the Covid-19 pandemic. At the time of data collection, about 8 in 10 reported doing so either all the time (36.6 per cent) or sometimes (46.0 per cent). Only a small percentage (3.9 per cent) had never worked remotely. Comparison with other New Zealand-based studies (see O'Kane et al., 2020; Stats NZ, 2020) is difficult due to differences in research design and the Covid-19 alert levels and restrictions in place when these studies were conducted. However, the empirical evidence from this paper suggests that, similar to these other studies, due to the forced flexibility imposed by the Covid-19 pandemic, remote working has since become commonplace in New Zealand's work environments.

More precisely, hybrid work, a flexible working arrangement where employees work partly remotely (mainly at home and/or another non-employer location) and partly in the physical workplace (Ostapenko et al., 2022) was at the time of the analysis the predominant model in New Zealand. According to our findings, about half of participants sometimes worked remotely from home. Similar trends are also seen in recent data from the United States (Parker, 2023), and Australia (Gallagher, 2021), with 41 per cent and 37.8 per cent, respectively. Due to the pandemic, it appears that most organisations now choose hybrid work arrangements (Gallagher, 2021); our results are consistent with this trend.

On the positive side, hybrid work offers increased flexibility, autonomy, and work-life balance (Gilson et al., 2022), but also brings challenges such as social isolation and the need for new skills and competencies (Griffiths, 2015). While the current study has only uncovered the prevalence of hybrid work in the New Zealand context, there are other aspects that future research will need to explore such as its impact on employees' productivity, recruitment and retention, and inclusion (Jaworski et al., 2023).

One notably finding was that the frequency of remote working was only associated with region of residence and not participants' gender and age. The result regarding region was expected as, during data collection, some parts of the country, particularly the Auckland region, were facing more socially restrictive Covid-19 measures due to community cases. Regarding gender, similar results have been reported by Abendroth et al. (2022) in Germany where no link between remote working and gender was found. The authors concluded that the pandemic in Germany reduced cultural barriers faced by women regarding remote working, resulting in no gender differences in the early months of the pandemic (Abendroth et al., 2022). Meanwhile, as no link was found between the frequency of working from home and age, further research is needed to compare with the findings of the current work and shed light on this topic using New Zealand-based evidence.

Another key contribution of this paper relates to the mixed results regarding the links of demographics with other key variables tested. it was found that age, but not gender, was statistically associated with preferences for working from home more often. While just over half the participants would like to work remotely from home more often, a much higher rate (66.7 per cent) was reported by those aged 18-29 years. Rates gradually decreased among older age groups. Studies on working from home and the pandemic suggest that younger employees have fewer opportunities to work remotely (Gallacher & Hossain, 2020) and tended to be more concerned about the impact of the pandemic on their careers compared to older employees (Ranta et al., 2020). These insights, thus, might provide possible explanations for the higher rates reported by younger participants in the present work.

The finding that preference for working more frequently from home was statistically independent of gender was also reported in previous studies of remote working during both conventional or business-as-usual times (Lim & Teo, 2000) and the Covid-19 pandemic (Brynjolfsson et al., 2020; Emanuel & Harrington, 2023; Nicks et al., 2021). This, however, differs from those of Iscan and Naktiyok (2005) and Nguyen and Armoogum (2021), who found that women tend to have a more positive view of remote working than men. Moreover, it is important to bear in mind that the finding in the current work is somewhat limited. When attitudes and preferences regarding remote working and gender are tested considering the marital/relationship and/or parenting status of employees, statistical differences are found (Dunatchik et al., 2021; Hartner-Tiefenthaler et al., 2022; Pennington et al., 2022). Therefore, we recommend that future research consider these specific aspects when looking at the relationship between gender and remote working in New Zealand.

Another contribution of the current work lies in the statistical association found between participants' perceptions regarding the impact of remote working on workplace culture and their preference towards working remotely from home more often. In this respect, seven in 10 participants who said the workplace culture was better due to flexible home working arrangements indicated that they would like to work from home more often. Meanwhile, over half of those who reported no change in the organisational culture indicated their preference for more frequent remote working. A lower but still substantial rate was reported among those who indicated that the work culture was worse (39.7 per cent).

Previous research points out that, to succeed, remote working requires a change in organisational culture (Bentley et al., 2016; Grant et al., 2018). A critical element for this is establishing a trusting culture inside the organisation as remote working depends on employee honesty and oversight from a distance (Grant et al., 2018). While challenging to achieve, trust can be enhanced through frequent communication and coaching (Grant et al., 2018) as well as effective team dynamics and organisational relationships (Bradley et al., 2023). Recently, a study of 40 American companies (Brown et al., 2021) reported that, in general, the organisational culture during the pandemic changed from a focus on high performance to emphasis on empathy, understanding, and mutual support. Meanwhile, Sull & Sull (2020) discovered a positive spike of perceived change in organisational culture after analysing 1.4 million employee reviews posted on Glassdoor, a platform that collects data and reviews from current and former employees about companies, salaries, and job openings. The reviews had been posted over the course of the previous five years. The surge was documented between April and August 2020, during the early stages of the Covid-19 epidemic in the United States. More recently, a qualitative study conducted in New Zealand by Bradley et al. (2023) came to the conclusion that employees' positive attitudes and experiences with remote working depend on an organisational culture that cultivates trust among socially connected teams with well-developed communication processes to support both relational cohesion and task completion. The current study's empirical findings add to the growing body of research relating organisational culture and the pandemic, since it was discovered that remote workers who have a positive impression of their company culture are more likely to appreciate working from home on a regular basis.

Future research on this topic is crucial, and using more thorough metrics will give a complete picture of how organisational culture affects remote workers. Meanwhile, the current work still offers a useful snapshot into the perceived impact of corporate culture on employees' views towards remote work. This finding is relevant for organisations thinking about redesigning and/or implementing flexible working arrangements.

The results also show that participants experienced diverse barriers to remote work. Among these issues faced, technology-related challenges such as slow internet connection emerged as the most prevalent barrier reported by participants who expressed a desire to work from home more frequently. This finding, while indicative, adds to current research interest and commentary about the impact of limited access to technology and quality of connectivity and its negative impact on effective communication, collaborative tasks, and training when working from home (Donati et al., 2021; Green et al., 2020). In addition, challenges related to the nature of the job (e.g., necessity to be physically present at the workplace for numerous face-to-face meetings) and environmental hindrances (e.g., inadequate space at home for remote work) were other commonly cited barriers by the respondents. Overall, these findings are consistent with the literature (see Green et al., 2020; O'Kane et al., 2020) about employees facing multifaceted challenges when working from home.

To conclude, this paper has provided a snapshot of working remotely from home in the context of the Covid-19 pandemic. The findings will inform organisations and policy makers considering the development and/or improvement of guidelines, and practices about retaining and/or assessing remote working and flexible/hybrid work arrangements. The experiences, preferences, and perceived barriers reported here by New Zealand employees about working from home provide the empirical evidence to inform those decisions. Also, organisations and employees are transitioning to a post-pandemic era with new business-as-usual circumstances. Thus, a natural progression of the current work is to examine the associations explored here longitudinally, for instance, using a two-wave panel design, to see whether attitudes have shifted. Moreover, the inclusion and careful assessment of other variables, such as ethnicity, marital status, type of employment, and digital access and skills, are suggested for future research to better understand the nuances of working remotely from home.

*Limitations*

A limitation is that the data were collected at one point in time, thus the findings represent a snapshot of participants' experiences. Furthermore, not only was data collection carried out under the challenging reality of the pandemic but also it took place when the country's regions were under different levels of the Covid-19 alert system. Thus, it is highly likely that this context has affected participants' self-reported work experiences and views of remote working. On the other hand, to understand how trends of remote working change over time, longitudinal evidence is needed. Another limitation is that the findings in this paper are limited by the original survey design. They represent a stocktake of current trends regarding remote working, therefore, a larger follow-up study will be needed.